\documentclass[12pt]{article}

\usepackage{latexsym}
\usepackage{amsmath}
\usepackage{multibox}
\usepackage{verbatim}
\usepackage{mathrsfs}

\usepackage{amssymb,amsmath}
\usepackage{amsthm}
\usepackage{amscd}
\usepackage{amssymb}
\usepackage{amsfonts}
\usepackage{url}
\usepackage{slashed}
\usepackage[latin1]{inputenc}
\usepackage[english]{babel}
\usepackage{putex}

 \csname
@addtoreset\endcsname{equation}{section}

\def\gz0{\gamma^{0}}

 \def\det{{\rm det\,}}

\def\scs#1{\section{\sc #1}}

\newcommand{\beq}{\begin{equation}}
\newcommand{\eeq}[1]{\label{#1}\end{equation}}
\newcommand{\bea}{\begin{eqnarray}}
\newcommand{\eea}[1]{\label{#1}\end{eqnarray}}

\def\bs{\begin{split}}
\def\es{\end{split}}
\def\ba{\begin{array}}
\def\ea{\end{array}}
\def\bec{\begin{center}}
\def\ec{\end{center}}
\def\ba{\begin{align}}
\def\ena{\end{align}}

\def\12{\frac{1}{2}}

\begin{document}

\begin{flushright}
{\today}
\end{flushright}

\vspace{25pt}

\begin{center}
%%%%%%%%%%%%%%%%%%%%%%%%%%%%%%%%%%%%%%%%%%%%%%%%%%%%%%%%%%%%%%%%%%%%
{\Large\sc Note on Gauge Invariance and Causal Propagation}\\
%%%%%%%%%%%%%%%%%%%%%%%%%%%%%%%%%%%%%%%%%%%%%%%%%%%%%%%%%%%%%%%%%%%%
\vspace{25pt}
{\sc Marc~Henneaux and Rakibur~Rahman}\\[15pt]
{\sl\small
Physique Th\'eorique et Math\'ematique \& International Solvay Institutes\\
Universit\'e Libre de Bruxelles, Campus Plaine C.P. 231, B-1050 Bruxelles, Belgium\\
\vspace{6pt}
e-mail: {\small \it
henneaux@ulb.ac.be, rakibur.rahman@ulb.ac.be}}\vspace{10pt}
%%%%%%%%%%%%%%%%%%%%%%%%%%%%%%%%%%%%%%%%%%%%%%%
\vspace{35pt}

{\sc\large Abstract}

\end{center}
%%%%%%%%%%%%%%%%%%%%%%%%%%%%%%%%%%%%%%%%%%%%%%%

Interactions of gauge-invariant systems are severely constrained by several consistency requirements.
One is the preservation of the number of gauge symmetries, another is causal propagation. For lower-spin
fields, the emphasis is usually put on gauge invariance that happens to be very selective by itself.
We demonstrate with an explicit example, however, that gauge invariance, albeit indispensable for
constructing interactions, may not suffice as a consistency condition. The chosen example that exhibits this
feature is the theory of a massless spin-3/2 field coupled to electromagnetism. We show that this system
admits an electromagnetic background in which the spin-3/2 gauge field may move faster than light. Requiring causal
propagation rules out otherwise allowed gauge-invariant couplings. This emphasizes the importance of causality
analysis as an independent test for a system of interacting gauge fields. We comment on the implications of
allowing new degrees of freedom and non-locality in a theory, on higher-derivative gravity and Vasiliev's
higher-spin theories.

\setcounter{page}{1}
\pagebreak

%%%%%%%%%%%%%%%%%%%%%%%%%%%%%%%%%%%
\scs{Introduction}\label{sec:intro}
%%%%%%%%%%%%%%%%%%%%%%%%%%%%%%%%%%%

Interacting theories of gauge fields are severely constrained. Powerful no-go theorems~\cite{No-go}
prohibit minimal coupling to gravity when the massless particle has spin $s\geq\tfrac{5}{2}$, as well
as to electromagnetism (EM) in flat space when $s\geq\tfrac{3}{2}$. Non-minimal couplings are still
allowed. In fact, trilinear vertices involving gauge fields of arbitrary spins can be classified by
using the light-cone formulation~\cite{Metsaev} and their covariant forms can be obtained by employing
either the Noether procedure~\cite{Karapet} or the BRST deformation scheme~\cite{BRST-BV,Nicolas,HLGR}.
These are higher-derivative interactions that result solely from the requirement of gauge invariance.
Is there anything that ensures that these terms respect causality and do not give rise to superluminal modes?

For low-spin systems the situation is somewhat different. In Yang-Mills theory, for example, potentially
bad interaction terms containing second time derivatives are eliminated by gauge symmetry itself~\cite{YM}.
Thus gauge invariance implies causality, which is nonetheless an independent consistency check.
For higher spins this is no longer true, so that requiring causality along with gauge invariance
becomes essential. The purpose of this letter is to highlight this point by proving it for some
specific example.

Indeed, gauge invariance does not suffice as a consistency condition. For massive higher-spin particles
one can introduce the St\"uckelberg fields to invent a fake gauge invariance, and then exploit this
symmetry to find deformations of the free theory~\cite{Zinoviev}. While this approach
enables us to find possible interactions for massive fields, it may leave the coupling constants
free. The requirement of causal propagation then fixes some, if not all, of these couplings~\cite{Buchbinder}.
The results of Ref.~\cite{Buchbinder} reaffirm, among others, the fact that the ``Velo-Zwanziger
acausality"~\cite{VZ} for a massive charged spin-2 particle in an EM background can be cured
not for an arbitrary magnetic dipole term as the gauge-invariant description might suggest, but precisely
when the gyromagnetic ratio is fixed to $g=2$~\cite{PRS}.

The organization of this letter is as follows. In Section~\ref{sec:Main} we take the free system of a massless
Rarita-Schwinger field and a photon, and consider its gauge deformations. All but one of the cubic couplings
are eliminated either by the lack of higher-order consistency or by the potential presence of propagating ghosts.
The only remaining vertex, however, leads to acausal propagation for the spin-$\tfrac{3}{2}$ field in a non-trivial
EM background, and we demonstrate this in Section~\ref{sec:Causality}. Finally, we make some remarks in
Section~\ref{sec:Conclusion} on the implications of adding new degrees of freedom and admitting non-locality in
a theory, and also on higher-derivative (super)gravity and Vasiliev's higher-spin theories as opposed to string theory.

\vskip 36pt

%%%%%%%%%%%%%%%%%%%%%%%%%%%%%%%%%%%%%%%%%%%%%%%%%%%%%%%%%%%%%%
\scs{The System of Spin-3/2 and Spin-1 Fields}\label{sec:Main}
%%%%%%%%%%%%%%%%%%%%%%%%%%%%%%%%%%%%%%%%%%%%%%%%%%%%%%%%%%%%%%

Let us consider the free theory containing a massless Rarita-Schwinger field $\psi_\mu$ and a photon $A_\mu$. It is
described by the action\footnote{We work in Minkowski space-time with mostly positive metric. The Clifford algebra is
$\{\gamma^\mu,\gamma^\nu\}=+2\eta^{\mu\nu}$, and $\gamma^{\mu\,\dagger}=\eta^{\mu\mu}\gamma^\mu$.
The Dirac adjoint is defined as $\bar{\psi}_\mu=\psi^\dagger_\mu\gamma^0$. The $D$-dimensional
Levi-Civita tensor, $\varepsilon_{\mu_1\mu_2\cdots\mu_D}$, is normalized as $\varepsilon_{01\cdots(D-1)}=+1$.
We define $\gamma^{\mu_1\cdots\mu_n}=\gamma^{[\mu_1}\gamma^{\mu_2}\cdots\gamma^{\mu_n]}$, where
the notation $[i_1\cdots i_n]$ means totally antisymmetric expression in all the indices $i_1,\cdots,i_n$
with a normalization factor $\tfrac{1}{n!}$.}
\beq \mathcal L_\text{free}=-i\bar{\psi}_\mu\gamma^{\mu\nu\rho}\partial_\nu\psi_\rho-\tfrac{1}{4}F_{\mu\nu}^2,\eeq{rs1}
which enjoys two abelian gauge invariances:
\beq \delta_\lambda A_\mu=\partial_\mu\lambda,\qquad\delta_\varepsilon\psi_\mu=\partial_\mu\varepsilon.\eeq{rs2}

To construct covariant cubic vertices one may employ, for example, the BRST deformation scheme for irreducible
gauge theories~\cite{BRST-BV}. The possible couplings are all non-minimal and may contain 1, 2 or 3
derivatives~\cite{Metsaev,HLGR,Taronna}. The parity-preserving covariant vertices are~\cite{HLGR}:
\beq \mathcal L_\text{cubic}=g_1\bar{\psi}_\mu\left(F^{\mu\nu}+\tfrac{1}{2}\gamma^{\mu\nu\rho\sigma}F_{\rho\sigma}\right)\psi_\nu
+g_2\left(\bar{\Psi}_{\mu\nu}\,\gamma^{\mu\nu\alpha\beta\lambda}\,\Psi_{\alpha\beta}\right)A_\lambda+g_3\bar{\Psi}_{\mu\alpha}
\Psi^\alpha_{~\nu}F^{\mu\nu},\eeq{cubic} where the coupling constants $g_i$'s are all real by Hermiticity. The EM and
spin-$\tfrac{3}{2}$ field strengths are respectively given by $F_{\mu\nu}=\partial_\mu A_\nu-\partial_\nu A_\mu$ and
$\Psi_{\mu\nu}=\partial_\mu\psi_\nu-\partial_\nu\psi_\mu$.

The 1-derivative Pauli term is a non-Abelian
deformation, i.e., it deforms the gauge algebra. The other two Abelian pieces do not deform the gauge transformations.
The 2-derivative vertex, which exists in $D\geq5$, is gauge invariant up to a total derivative, while the 3-derivative
one is just a 3-curvature term (Born-Infeld type). The non-Abelian piece faces obstruction in a local theory
beyond the cubic order~\cite{HLGR}. In other words, if $g_1\neq0$, the cubic couplings~(\ref{cubic}) are killed
by the quartic-order consistency unless one adds new degrees of freedom and/or admits non-locality.
If we demand locality, the original system~(\ref{rs1}) of a spin-$\tfrac{3}{2}$ gauge field and a photon has consistent
deformation up to all orders if and only if \beq g_1=0.\eeq{g1out} Given this, the spin-$\tfrac{3}{2}$ equations of
motion (EoM) become
\beq \gamma^{\mu\nu\rho}\partial_\nu\psi_\rho+ig_2\gamma^{\mu\nu\rho\alpha\beta}F_{\alpha\beta}\Psi_{\nu\rho}
-2ig_3\partial_\nu\left(F^{\alpha[\mu}\Psi^{\nu]}_{~\alpha}\right)=0.\eeq{EoM3/20}
These EoMs necessarily contain second time derivatives for generic $F_{\mu\nu}$; this is due to the presence of the last
term in Eq.~(\ref{EoM3/20}), which can be made explicit by writing
\beq -2\partial_\nu\left(F^{\alpha[\mu}\Psi^{\nu]}_{~\alpha}\right)=F^{\mu\nu}\left(\Box
\psi_\nu-\partial_\nu\partial\cdot\psi\right)-F^{\alpha\beta}\partial^\mu\partial_\alpha\psi_\beta+\text{lower time derivatives}.
\eeq{EoM3/21} As a result, the system may have propagating ghosts since the gauge symmetry no longer guarantees the removal of all
but the physical polarizations. The remedy is simply to remove the 3-derivative coupling, i.e., to set \beq g_3=0.\eeq{removeghost}
With the only non-zero coupling constant $g_2\equiv g$, the EoMs now reduce to
\beq \left[\gamma^{\mu\nu\rho}+2ig\gamma^{\mu\nu\rho\alpha\beta}F_{\alpha\beta}\right]\partial_\nu\psi_\rho=0.\eeq{EoM3/22}
On the other hand, the spin-1 field obeys the EoMs \beq \partial_\mu F^{\mu\nu}=J^\nu\,,\eeq{EoM1}
where the current $J^\mu$ comprises some spin-$\tfrac{3}{2}$ bilinears.
Below we will study a possible solution of the system of equations~(\ref{EoM3/22}) and~(\ref{EoM1}).

\vskip 36pt

%%%%%%%%%%%%%%%%%%%%%%%%%%%%%%%%%%%%%%%%%%%%%%
\scs{Causality Analysis}\label{sec:Causality}
%%%%%%%%%%%%%%%%%%%%%%%%%%%%%%%%%%%%%%%%%%%%%%

Let us consider small fluctuations of the spin-$\tfrac{3}{2}$ field. The right-hand side of Eq.~(\ref{EoM1}) can therefore be neglected,
so that the photon EoMs have the solution \beq F_{\mu\nu}=\text{constant}.\eeq{sol1} In this EM background, we would like to investigate the
propagation of the spin-$\tfrac{3}{2}$ field as a probe. Its dynamics is governed by the Lagrangian equation~(\ref{EoM3/22}), which has
the same number of components as the vector-spinor $\psi_\mu$, i.e., $D\times2^{[D]/2}$ components in $D$ space-time dimensions, with
$[D]\equiv D+\tfrac{1}{2}\left[(-1)^D-1\right]$. Now the $\mu=0$ component of Eq.~(\ref{EoM3/22}) does not
contain any time derivative and hence constitutes a constraint, which renders $2^{[D]/2}$ of the components non-dynamical.
Because $\dot\psi_0$ never appears in Eq.~(\ref{EoM3/22}), $\psi_0$ is just a Lagrange multiplier, and thus one gets rid of
additional $2^{[D]/2}$ components. Finally, one can do a complete gauge fixing by setting, for example\footnote{Here
$i=1,2,...,D-1$ corresponds to the spatial components. To see that
this is indeed a complete gauge fixing, suppose this is not the case. Then the residual gauge parameter must satisfy the
constraint $\gamma^i\partial_i\varepsilon=0$, and hence, in particular,  the Laplace equation: $\nabla^2\varepsilon=0$.
Given that the gauge parameter should vanish at spatial infinity, the only possible solution is $\varepsilon=0$. Therefore,
there is no residual gauge symmetry.}
\beq \gamma^i\psi_i=0,\eeq{gf} to end up having a correct total of $(D-3)\times2^{[D]/2}$ propagating degrees of freedom (fields
and momenta) for a massless spin-$\tfrac{3}{2}$ field in $D$ dimensions.

In order to see if the propagation of the physical components is inside the light cone, we take recourse of the shock-wave
formalism~\cite{SK}. The method relies on the fact that characteristic surfaces for wave propagation are those that support
discontinuities in the highest-order derivative terms in the EoMs. Let us denote the discontinuity across the characteristic
as \beq [\partial_\mu\psi_\nu]=\zeta_\mu\tilde\psi_\nu,\eeq{discon} where $\zeta_\mu$ is a vector normal to the characteristic
surface and $\tilde\psi_\mu$ is some vector-spinor defined on the same. Thus Eq.~(\ref{EoM3/22}) yields
\beq \left[\gamma^{\mu\nu\rho}+2ig\gamma^{\mu\nu\rho\alpha\beta}F_{\alpha\beta}\right]\zeta_\nu\tilde\psi_\rho=0.\eeq{MasterEqn}
If the wave propagation is causal, any component of $\tilde\psi_\mu$ must vanish for a time-like $\zeta_\mu$ unless it corresponds
to an unphysical or a non-dynamical mode. Without loss of generality, let us choose the time-like vector $\zeta_\mu=(1,0,...,0)$.
While the $\mu=0$ component of Eq.~(\ref{MasterEqn}) is trivially satisfied, the space-like components give
\beq \left[\gamma^{ij}+2ig\gamma^{ijkl}F_{kl}\right]\tilde\psi_j=0.\eeq{psikfirst}
On the other hand, the discontinuity of the time derivative of the gauge choice~(\ref{gf}) across the characteristic sets
\beq \gamma^i\tilde\psi_i=0.\eeq{0gone} Let us first implement this consequence of the gauge choice in Eq.~(\ref{psikfirst})
to write \beq \left[\mathbf{1}\delta^{ij}-2ig\,\gamma^{ijkl}F_{kl}\right]\tilde\psi_j=0.\eeq{psik}
If $g$ vanishes, clearly all $\tilde\psi_i=0$, so that the wave propagation is causal as expected.\footnote{Because $\tilde\psi_0$
corresponds to a Lagrange multiplier, not a dynamical field, it is irrelevant for our discussion.} When $g\neq0$,
we would like to see if Eq.~(\ref{psik}) could admit non-trivial solutions for $\tilde\psi_i$.

For the rest of the analysis, let us consider $D=5$. The components of the electric field $\vec{E}$ and the magnetic field $\mathbb{B}$,
which is an antisymmetric rank-2 spatial tensor, are given by \beq F^{0i}=E^i,\qquad F^{ij}=\tfrac{1}{2}\varepsilon^{ijkl}B_{kl}.\eeq{EBfield}
In $D=5$, there are two independent EM field invariants: \beq \text{Tr}F^2=2\vec{E}^2+\text{tr}\mathbb{B}^2,\qquad \text{Tr}F^4
=2(\vec{E}^2)^2+\text{tr}\mathbb{B}^4-(\vec{E}\times\mathbb{B})^2,\eeq{Invariants} where $\text{Tr}$ and $\text{tr}$ denote traces
in 5-dimensional Minkowski space-time and 4-dimensional space respectively, and $\vec{E}\times\mathbb{B}$ is the spatial vector with the
$i$-th component $\varepsilon^{ijkl}E_jB_{kl}$. With a nonzero $\mathbb{B}$ field, one finds that $\text{tr}\mathbb{B}^2$ is always negative
whereas $\text{tr}\mathbb{B}^4$ is always positive.

Let us now consider Eq.~(\ref{psik}) to see if there exist non-zero solutions for $\tilde{\psi}_\mu$. These equations
involve only the magnetic field, which we can choose to be
\beq
\mathbb{B}=\left(
\begin{aligned}
\begin{tabular}{c c|c c}
  $0$ & $B_{12}$ & $0$ & $0$\\
  $-B_{12}$ & $0$ & $0$ & $0$\\
  \hline
  $0$ & $0$ & $0$ & $B_{34}$\\
  $0$ & $0$ & $-B_{34}$ & $0$\\
\end{tabular}
\end{aligned}
\right),
\eeq{Magnetic}
thanks to the spatial-isotropy-preserving gauge choice~(\ref{gf}). For simplicity, we stick to the special case
$B_{12}=B\neq0$, and $B_{34}=0$. The only non-zero components of the EM field strength are then
$F_{34}=-F_{43}=B$. In this case, Eq.~(\ref{psik}) reduces to
\beq
\left(
\begin{aligned}
\begin{tabular}{c c|c c}
  $\mathbf{1}$ & $-i\mathbf\Gamma$ & $\mathbf 0$ & $\mathbf 0$\\
  $i\mathbf\Gamma$ & $\mathbf{1}$ & $\mathbf 0$ & $\mathbf 0$\\
  \hline
  $\mathbf 0$ & $\mathbf 0$ & $\mathbf 1$ & $\mathbf 0$\\
  $\mathbf 0$ & $\mathbf 0$ & $\mathbf 0$ & $\mathbf 1$\\
\end{tabular}
\end{aligned}
\right)\left(
\begin{aligned}
\begin{tabular}{c}
  $\tilde\psi_1$\\
  $\tilde\psi_2$\\
  $\tilde\psi_3$\\
  $\tilde\psi_4$\\
\end{tabular}
\end{aligned}
\right)=\left(
\begin{aligned}
\begin{tabular}{c}
  $\underline{0}$\\
  $\underline{0}$\\
  $\underline{0}$\\
  $\underline{0}$\\
\end{tabular}
\end{aligned}
\right),
\eeq{last}
where $\mathbf\Gamma=4gB\,\gamma^{1234}$.
From the block diagonal form of the matrix it is clear that \beq \tilde\psi_3=0,\qquad \tilde\psi_4=0.\eeq{34out}
Then the consequence~(\ref{0gone}) of the gauge choice~(\ref{gf}) reduces to
\beq \gamma^1\tilde\psi_1+\gamma^2\tilde\psi_2=0.\eeq{12related} This enables us to write
$\tilde\psi_2=\gamma^{12}\tilde\psi_1$, so that Eq.~(\ref{last}) gives
\beq \left(\mathbf{1}+4igB\,\gamma^{34}\right)\tilde\psi_1=0.\eeq{good}
The determinant of this coefficient matrix is \beq \det\left(\mathbf{1}+4igB\,\gamma^{34}\right)=\left[\left(4gB\right)^2-1\right]^2.
\eeq{good1} If $g\neq0$, this will vanish if the magnetic field is \beq g^2B^2=\frac{1}{16}\,.\eeq{bad}
Thus, indeed $\tilde\psi_{1,2}$ may have non-trivial solutions.

Note that we are interested only in small values of the EM field invariants:
\beq g^2|\text{Tr}F^2|\ll1,\qquad  g^4|\text{Tr}F^4|\ll1.\eeq{SmallInvariants} Otherwise,
various instabilities appear~\cite{instability} and the concept of long-lived propagating particles ceases to make sense.
But the non-trivial solutions for $\tilde\psi_\mu$ show up even for infinitesimally small values of the EM
field invariants if the electric field $\vec{E}$ is such that \beq \vec E^2=-\tfrac{1}{2}\text{tr}\mathbb{B}^2+\epsilon_1,\qquad
(\vec E\times\mathbb{B})^2=\tfrac{1}{2}(\text{tr}\mathbb{B}^2)^2+\text{tr}\mathbb{B}^4+\epsilon_2,\eeq{chooseE} where
$|\epsilon_1|\ll1/g^2$ and $|\epsilon_2|\ll1/g^4$.
We conclude that superluminal propagation takes place within the regime of physical interest. To cure this pathology,
one must set \beq g\equiv g_2=0.\eeq{gzero} Therefore, causal propagation in the absence of additional degrees of freedom
admits no cubic couplings at all, although they are allowed by gauge invariance alone.

\vskip 36pt
%%%%%%%%%%%%%%%%%%%%%%%%%%%%%%%%%%%
\scs{Remarks}\label{sec:Conclusion}
%%%%%%%%%%%%%%%%%%%%%%%%%%%%%%%%%%%

In this paper, we presented an explicit example to make the point that gauge invariance alone does not guarantee the
consistency of a theory of interacting gauge fields and that causality must be added as an independent requirement.
To the best of our knowledge, this is the first time such an analysis has been done for massless fields.

As pointed out in Ref.~\cite{HLGR}, the $g_2$ term analyzed here is of Chern-Simons (C-S) type. Now, for bosonic fields
with free EoMs containing second-order derivatives, a C-S term cannot lead to shock-wave acausal behavior because its
contribution to the EoMs contains only first-order derivatives. The reason why the $g_2$ term plays a critical role here
is that the free fermionic EoMs are first order in derivatives, and so the C-S term competes with the unperturbed Lagrangian
in the shock-wave causality analysis.

The three cubic couplings appearing in Eq.~(\ref{cubic}) were considered piecemeal in the subsequent discussion and it
is legitimate to do so from a gauge-theoretic point of view. Note that the addition of a dynamical graviton
in the theory removes the obstruction of the non-Abelian piece at higher orders while keeping locality intact. Indeed,
this vertex is present in $\mathcal N=2$ gauged supergravity~\cite{fpvn}. Decoupling gravity by taking
$M_\text{P}\rightarrow\infty$ kills this Pauli term because the dimensionful coupling constant goes like
$1/M_\text{P}$~\cite{fpvn}. Given this, one may require that the consistency of the gauge deformation itself
should define the degrees of freedom to begin with. With the inclusion of gravity, all the pieces in the
vertex~(\ref{cubic}) may pass the quartic-order consistency to give rise to some higher-derivative counterpart
of $\mathcal N=2$ supergravity. One would like to know if this higher-derivative supergravity theory is causal.
On the other hand, integrating out the massless graviton gives a non-local theory of spin-$\tfrac{3}{2}$
and spin-1 fields with the Pauli term. Thus one can forgo locality for the sake of higher-order consistency.
In the resulting theory, however, the issue of causality becomes very obscure.

Similarly, one can take ghost-free higher-derivative gravity theories, i.e., Lanczos-Lovelock gravities~\cite{Lovelock},
to see if they suffer from superluminality. A canonical analysis of this is intricate~\cite{TZ}, but the existence
of such a pathology may not come as a surprise. After all, AdS/CFT analysis shows that generic values of the
higher-derivative couplings in the bulk afflict the boundary theory with superluminal modes~\cite{Lovelockbad}.
While such higher-derivative terms do exist in the $\alpha'$ expansion of string theory,\footnote{The smallness of $\alpha'$
saves the day because ``small" couplings are still consistent~\cite{Lovelockbad}.} they show up with a plethora of other
terms; in the end, the theory contains infinitely many derivatives to become essentially non-local. Ghost-free nonlinear
theories also exist for massive gravity~\cite{Gabadadze}, but they are plagued by superluminal propagation~\cite{Andrew}.
The lack of contradiction between ghostlessness and acausality in this case is very similar in spirit to the
fact that gauge invariance for massless models and causality are independent requirements, which must be
checked separately.\footnote{We thank S.~Deser for comments stressing this point.}

Neither does gauge invariance imply the consistency of the S matrix. This issue is addressed
in~\cite{mirian}, and the analysis uses cubic couplings in Minkowski space, just like ours does.
The conclusion of~\cite{mirian} is that essentially all local higher-spin cubic vertices are ruled out
in flat space, even with an infinite tower of fields. Higher-spin gauge theories on a flat background may
still make sense if they incorporate extended and possibly non-local objects, like the stringy
Pomerons~\cite{Pomeron}. It remains to be seen how causality works in this case.

In view of our results, it would be interesting to see whether Vasiliev's higher-spin theories~\cite{Vasiliev} pass
the test of causal propagation. One might, however, argue that in these theories the metric itself and hence the
light cone have no gauge-invariant meaning. Therefore, the issue of causality becomes tricky. Yet if the (infinite
tower of) higher-spin excitations are treated as perturbations, it would still make sense to ask if they travel
within the light cone. Requiring causality for Vasiliev's theories may shed important light on string theory by
telling us whether consistent gauge theories for higher-spin particles necessarily call for the full tower of string
states in the tensionless limit.

\vskip 36pt

%%%%%%%%%%%%%%%%%%%%%%%%%%
\section*{Acknowledgments}
%%%%%%%%%%%%%%%%%%%%%%%%%%

We would like to thank I.~L.~Buchbinder for questions that prompted this work and A.~Campoleoni, M.~Porrati and M.~Vasiliev
for useful comments. MH gratefully acknowledges support from the Alexander von Humboldt Foundation through a Humboldt
Research Award and support from the ERC through the ``SyDuGraM" Advanced Grant. RR is a Postdoctoral Fellow of the
Fonds de la Recherche Scientifique-FNRS. His work is partially supported by IISN-Belgium (conventions 4.4511.06 and
4.4514.08), by the ``Communaut\'e Fran\c{c}aise de Belgique" through the ARC program and by the ERC Advanced Grant ``SyDuGraM."

\end{document}